\documentclass[useAMS,usenatbib]{mn2e}

\usepackage{float} 
\usepackage{epsfig}
\usepackage{graphicx}
\usepackage[latin1]{inputenc}
\usepackage{latexsym}
\def\simlt{\ \raise -2.truept\hbox{\rlap{\hbox{$\sim$}}\raise5.truept   %
\hbox{$<$}\ }}
\def\simgt{\ \raise -2.truept\hbox{\rlap{\hbox{$\sim$}}\raise5.truept   %
\hbox{$>$}\ }}                                                          %

\def\be{\begin{equation}}
\def\ee{\end{equation}}
\def\newline{\hfil\break}

\def\la{\mathrel{\hbox{\rlap{\hbox{\lower4pt\hbox{$\sim$}}}\hbox{$<$}}}}
\def\ga{\mathrel{\hbox{\rlap{\hbox{\lower4pt\hbox{$\sim$}}}\hbox{$>$}}}}

\title[Reacceleration of DM-produced electrons]{Effect of turbulent reacceleration on electrons produced by dark matter annihilation in the Coma cluster}
\author[P. Marchegiani]{P. Marchegiani$^{1}$\thanks{E-mail: Paolo.Marchegiani@wits.ac.za}\\
$^{1}$School of Physics, University of the Witwatersrand, Private Bag 3, 2050-Johannesburg, South Africa\\
}

\begin{document}

\date{Accepted: 2019 June 28. Received: 2019 June 28; in original form: 2019 March 12}

\pagerange{\pageref{firstpage}--\pageref{lastpage}} \pubyear{2019}

\maketitle

\label{firstpage}

\begin{abstract}
In this paper we study the effect of reacceleration provided by turbulences on electrons produced by dark matter (DM) annihilation in the Coma cluster. We use a simplified phenomenological model to describe the effect of the turbulences, and explore a limited subset of three possible DM models for neutralino particles with different mass and annihilation channel. We find that, for values of the annihilation cross section of the order of the upper limits found with Fermi-LAT measurements in astrophysical objects, and for conservative values of the boosting factor due to DM substructures, the reacceleration due to turbulences can enhance the radio emission produced by DM-originated electrons up to the level of the observed flux of the radio halo in Coma, for moderate reacceleration intensity in relatively short times. Therefore we conclude that, even if 
it is not possible to distinguish between the fits obtained in this paper because of the scattering present in the radio flux data,
the electrons produced by DM annihilation can be possible seed electrons for the reacceleration, as well as secondary electrons of hadronic origin. A possible discriminant between these two classes of models is the flux produced in the gamma ray band, that in the case of DM-originated electrons should be more than two orders of magnitude smaller than the present Fermi-LAT upper limits, whereas in the hadronic case the expected gamma ray flux should be close to the value of present upper limits. 
\end{abstract}

\begin{keywords}
galaxies: clusters: general - dark matter.
\end{keywords}


\section{Introduction}     

Electrons produced in dark matter (DM) annihilation processes in galaxy clusters can, in principle, produce a diffuse radio emission when interacting with the intra-cluster magnetic field (Colafrancesco, Profumo \& Ullio 2006). The DM-originated radio emission can be compared with the diffuse radio emission observed in a number of clusters in the form of giant radio halos (e.g. Feretti et al. 2012, Brunetti \& Jones 2014, van Weeren et al. 2019).

Previous studies have shown that, while in some clusters like the Bullet cluster (Marchegiani \& Colafrancesco 2015) and A520 (Marchegiani, Colafrancesco \& Khanye 2019) the DM-originated radio emission can be important at most in some specific regions of the cluster, in the Coma cluster the whole flux of the radio halo can be produced by DM annihilation without violating the Fermi-LAT gamma ray upper limits (Marchegiani \& Colafrancesco 2016).

However, this last result was based on quite optimistic assumptions on the values of the DM annihilation cross section and of the boosting factor due to the presence of DM substructures. In fact, the values of the annihilation cross section were taken from a study of the gamma ray excess in the Galactic centre (Abazajian \& Keeley 2016), while other studies of the Galactic centre excess itself (e.g. Calore et al. 2015) or the gamma ray upper limits in dwarf galaxies (Ackermann et al. 2015) have provided lower values for the annihilation cross sections. Also, it is possible that the Galactic centre excess is not due to DM annihilation but to astrophysical sources (e.g. Bartels et al. 2018), and in this case the value of the annihilation cross section determined from the Galactic centre excess should be reduced.

Another strong argument against the interpretation of the origin the radio halos in galaxy clusters in terms of DM annihilation is that, while the amount of the electrons and consquently the intensity of the produced radio emission in DM models should strongly correlate only with the mass of the cluster, radio halos are usually hosted in clusters in a disturbed state, indicating an ongoing or recent merging event. Moreover, there is a bimodality between disturbed and relaxed clusters with similar mass and X-ray luminosity, with the first ones often hosting bright radio halos, and the second ones having radio upper limits one or two orders of magnitude smaller than the luminosities observed in disturbed clusters (e.g. Cassano et al. 2013). The luminosity of radio halos in fact results to be correlated, other than with the mass or the X-ray luminosity of the cluster, also with the level of disturbance of the cluster itself (Yuan, Han \& Wen 2015, Eckert et al. 2017).

These evidences have led to consider the magnetohydrodynamics (MHD) turbulences that develop during a merging event as a possible source of diffuse reacceleration of seed relativistic electrons, that can be of primary origin, resulting for example from shock acceleration or AGN injection (e.g. Brunetti et al. 2001), or of secondary origin, produced in hadronic interactions between cosmic ray protons and thermal nuclei (e.g. Brunetti \& Blasi 2005). At the moment the origin of the seed electrons is still not well determined (e.g. Pinzke, Oh \& Pfrommer 2017).

The effect of turbulent reacceleration should be effective also on electrons produced in DM annihilation processes. This effect in principle can increase the radio emission due to DM-originated electrons, and explain the correlation between the presence of a radio halo and the level of disturbance of a cluster. Until now the effect of turbulent reacceleration has not been considered in DM models.

In this paper, we perform a first study of the effect of turbulent reacceleration on DM-produced electrons, taking the Coma cluster as a case of study. We apply a simple phenomenological model of turbulent reacceleration, that does not consider the details of the properties of the MHD waves produced by the turbulences, to the electrons produced by DM annihilation for a small set of DM models, for neutralinos with three representative mass values and annihilation channels, using conservative values for the annihilation cross section and the substructures boosting factor, and we study if the resulting radio emission has an intensity and a spectrum similar to the ones observed in the Coma radio halo. A full model, considering the details of the reacceleration process, a larger number of DM models, and taking into account also the spatial properties of the radio emission in Coma and other clusters, is beyond the goals of this paper, and will be considered in the future.

Throughout the paper, we use a flat, vacuum--dominated cosmological model following the results of the Planck satellite, with $\Omega_m = 0.308$, $\Omega_{\Lambda} = 0.692$ and $H_0 =67.8$ km s$^{-1}$ Mpc$^{-1}$ (Ade et al. 2016). With these values the luminosity distance of Coma at $z=0.023$ is  $D_L=104$ Mpc, and 1 arcmin corresponds to 28.9 kpc.

\section{Methods}

The time evolution of the electrons spectrum, as resulting from the reacceleration produced by turbulences and the energy losses, and in presence of a steady source as in the case of the DM annihilation, is given by the equation:
\begin{eqnarray}
\frac{\partial n(p)}{\partial t} & = & \frac{\partial}{\partial p} \left[ \left(-\frac{2}{p}D_{pp}+\sum_i \left|\frac{dp}{dt}\right|_i \right)n(p) \right. \nonumber\\
 & & \left. + D_{pp}\frac{\partial n(p)}{\partial p}\right] +Q(p)
 \label{eq.diffusion}
\end{eqnarray}
(e.g. Schlickeiser 2002), where $D_{pp}$ is the diffusion coefficient in the momentum space, $|dp/dt|_i$ contains the effect of the energy losses due to the interaction with the magnetic field, the CMB photons, and the thermal gas (e.g. Sarazin 1999), and $Q(p)$ is the source term due to DM annihilation.

Since the usual form of the diffusion coefficient as resulting from different properties of the MHD waves associated to the turbulences is of the type $D_{pp}\propto p^2$ (Brunetti \& Lazarian 2007, 2011), we assume a simple phenomenological form as
\begin{equation}
D_{pp}= \frac{\chi}{2} p^2 ,
\end{equation} 
so that in eq.(\ref{eq.diffusion}) we can write $-2D_{pp}/p=-\chi p$, and the corresponding characteristic acceleration time is
\begin{equation}
\tau_{acc}=\frac{p^2}{4D_{pp}}=\frac{1}{2\chi}
\end{equation}
(Brunetti \& Lazarian 2007). In this way, the strength of the reacceleration is expressed through the parameter $\chi$, that, for typical characteristic acceleration times of the order of several hundreds of Myr, takes values of the order of $10^{-17}-10^{-16}$ s$^{-1}$, and results to be higher in modulus than the loss term for Lorentz factors of the electrons of the order of $10^3-10^4$, i.e. in the spectral region of radio-emitting electrons (e.g. Brunetti et al. 2001).

In order to simplify the calculations, we adopt a value of $\chi$ not depending on the position inside the cluster, and we do not consider the effect of the spatial diffusion of the electrons, that in DM models can be important in modifying the radial shape of the resulting surface brightness emission close to the central cusp of the DM distribution (e.g. Colafrancesco et al. 2006). For the energy losses, we instead consider the radial profile of the thermal gas $n_{th}(r)$ as derived from X-ray measures (Briel, Henry \& Boehringer 1992), i.e. a beta model
\begin{equation}
n_{th}(r)=n_{th,0} \left[1+\left(\frac{r}{r_c}\right)^2\right]^{-3\beta/2}
\end{equation}
(Cavaliere \& Fusco-Femiano 1976) with $n_{th,0}=3.4\times10^{-3}$ cm$^{-3}$, $\beta=0.75$, and $r_c=0.3$ Mpc, and a magnetic field as derived from Faraday Rotation measures, with $B(r)\propto n_{th}(r)^{1/2}$ and central value $B_0=4.7$ $\mu$G (Bonafede et al. 2010). As a consequence of neglecting spatial diffusion, eq.(\ref{eq.diffusion}) can be solved separately at each position in the cluster. Consistently, we 
focus our discussion mainly on the radio flux integrated over the cluster volume; 
for sake of completeness we will show also the azimuthally averaged profile of the radio surface brightness in a few cases, without trying to use this further information to constrain the DM models in this paper.

Compared to the case of reacceleration of electrons produced in hadronic interactions (e.g. Brunetti \& Blasi 2005), the case of electrons produced by DM annihilation is simplified because MHD waves associated to turbulences are not expected to act on DM particles; for this reason, eq.(\ref{eq.diffusion}) can be directly used with a source spectrum $Q(p)$ that is not affected by the effect of the turbulences. The source spectrum for DM annihilation is given by:
\begin{equation}
Q (E,r) = {\cal B} \langle \sigma v\rangle \frac{dN(E)}{dE}\mathcal{N}_{DM} (r)
\label{source.term}
\end{equation}
(Colafrancesco et al. 2006), where $\langle \sigma v\rangle$ is the thermally-averaged annihilation cross-section, $dN/dE$ is the spectrum of electrons production that can be calculated using the DarkSusy package (Gondolo et al. 2004), and ${\cal B}$ is a multiplicative boosting factor produced by the effect of DM substructures (Pieri et al. 2011). The function  $\mathcal{N}_{DM} (r) $ is the DM pair density, $\mathcal{N}_{DM} (r) = (\rho (r) )^2/(2 M_{DM}^2)$, where $M_{DM}$ is the mass of the DM particle, and $\rho (r)$ is the DM density profile, that we assume to have a Navarro, Frenk \& White (1996) form with parameters values related to the mass of the DM halo (Bullock et al. 2001). In the Coma cluster, for a total mass of $1.24\times10^{15}$ M$_\odot$ (Okabe et al. 2014), the derived DM profile is
\begin{equation}
\rho(r)=\frac{\rho_s}{\left(\frac{r}{r_s}\right)\left(1+\frac{r}{r_s}\right)^2},
\label{dens.dm}
\end{equation}
with $\rho_s=4.83\times10^3\rho_c$, being $\rho_c$ the critical density of the Universe, and $r_s=0.545$ Mpc.

In the following we will use the neutralino as a candidate DM particle, referring to three specific models with different representative DM particle mass and annihilation final state (e.g. Colafrancesco et al. 2011): a model with $M_{DM}=9$ GeV and annihilation final state $\tau^+\tau^-$, a model with $M_{DM}=60$ GeV and annihilation final state $b \bar b$, and a model with $M_{DM}=500$ GeV and annihilation final state $W^+W^-$.

We will use values of the annihilation cross section corresponding to the upper limits found in dwarf galaxies by the Fermi-LAT Collaboration (Ackermann et al. 2015); this assumption implies that the results that will be obtained are in accordance with the observational upper limits, but that need to be considered as upper limits too; the real signal produced by DM annihilation in fact might be smaller also by many orders of magnitude. 

We will also use a conservative value of the substructures boosting factor of ${\cal B}=30$; the value of the boosting factor is in fact still not well constrained, because it strongly depends on the value of the minimum substructures mass, that at the moment is well below the resolution of cosmological simulation (e.g. Ando \& Nagai 2012). The common assumption of a minimum mass of the order of the WIMP free streaming scale, $\sim10^{-6}$ M$_\odot$ (Green, Hofmann \& Schwarz 2005), can provide high values of the boosting factor, up to 700 or more (Springel et al. 2008, Gao et al. 2012), intermediate values of the order of 30--40 (Anderhalden \& Diemand 2013, Sanchez-Conde \& Prada 2014), or smaller values of the order of 2 (Storm et al. 2013, 2017), depending on the details of the cosmological simulations that have been used; other effects can also increase the value of the boosting factor, as the DM contraction due to baryon dissipation, that can provide a value of 2--6 (Ando \& Nagai 2012), a steeper cusp profile in the inner part of sub-halos, that can produce a boosting factor of the order of 70 (Ishiyama 2014), or the effect of tidal stripping, that can increase the boosting factor by 2--5 (Bartels \& Ando 2015); for these reasons, we consider a value of 30 as a quite conservative value for the boosting factor.

Following previous studies of the reacceleration of secondary electrons of hadronic origin (e.g. Brunetti et al. 2012), we adopt a process in two steps: in the first step we calculate the equilibrium spectrum in eq.(\ref{eq.diffusion}) without reacceleration (i.e. for $D_{pp}=0$), representing the phase of the cluster life when reacceleration is not effective, and in the second step we assume a constant non-zero value of $D_{pp}$ and, taking the equilibrium spectrum previously calculated as the initial spectrum, and maintaning the same source spectrum due to DM annihilation, we calculate the evolution with time of the electrons spectrum until a maximum value of the acceleration period $T_{acc}$, that is supposed to be at maximum of the order of 1 Gyr before turbulences are dissipated (e.g. Brunetti \& Lazarian 2007). For the resulting electrons spectrum, we calculate the synchrotron emission for the magnetic field derived from Farady Rotation measures (Bonafede et al. 2010), and compare the result with the spectrum of the radio halo observed in the Coma cluster (Thierbach, Klein \& Wielebinsk 2003 and references therein).

In our calculations, eq.(\ref{eq.diffusion}) is solved adopting a numerical approach based to a finite difference scheme applied to a Fokker-Planck equation (Chang \& Cooper 1970, Park \& Petrosian 1996, Donnert \& Brunetti 2014).

\section{Results}

We consider in our calculations two possible values of the parameter $\chi$, that is associated to the intensity of the reacceleration provided by the turbulences: $\chi=1\times10^{-16}$ s$^{-1}$ and $\chi=5\times10^{-17}$ s$^{-1}$, with the first one being associated to a stronger level of turbulence. These values correspond to characteristic acceleration times of the order of $\tau_{acc}\sim160-320$ Myr, and therefore are associated to a level of turbulence comparable to what is usually assumed in models for reacceleration of electrons of hadronic origin, that is of the order of $\tau_{acc}\sim200-300$ Myr (e.g. Brunetti, Zimmer \& Zandanel 2017). We note that, while in hadronic models, where the equilibrium and the source spectra have a power-law shape for electrons emitting in the radio band, it is possible to constrain the value of $\tau_{acc}$ from the frequency where the radio spectrum shows a steepening, the electrons spectra due to DM annihilation present an intrinsic steepening at high energy due to the fall of the production spectrum in correspondance of the DM mass, and therefore constraining the value of $\tau_{acc}$ on the basis of the frequency steepening is less immediate.

\begin{figure}
\centering
\begin{tabular}{c}
\includegraphics[width=\columnwidth]{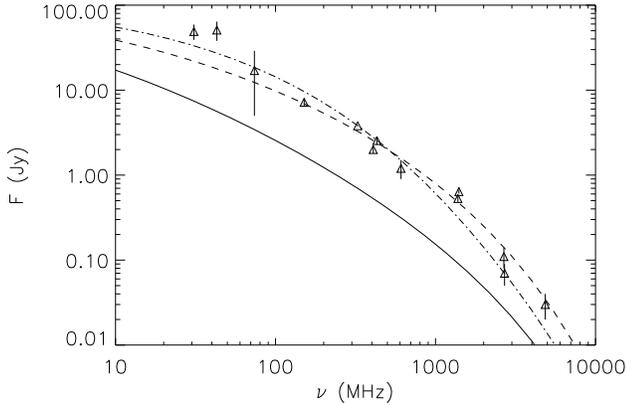}
\end{tabular}
\caption{Spectrum of the radio emission produced for a DM model with mass 9 GeV, annihilation final state $\tau^+\tau^-$, annihilation cross section $\langle \sigma v\rangle=4\times10^{-27}$ cm$^3$ s$^{-1}$, and boosting factor ${\cal B}=30$ for the equilibrium case without reacceleration (solid line), the case with reacceleration with $\chi=1\times10^{-16}$ s$^{-1}$ after $T_{acc}=1.2\times10^8$ yr (dashed line), and the case with reacceleration with $\chi=5\times10^{-17}$ s$^{-1}$ after $T_{acc}=4.5\times10^8$ yr (dot-dashed line).}
\label{radio_9tt}
\end{figure}

In Figure \ref{radio_9tt} we show the radio spectrum produced in the case of a neutralino with mass 9 GeV and annihilation final state $\tau^+\tau^-$ for the equlibrium case without reacceleration, and the spectra produced for the two values of $\chi$ we have used, choosing the value of the total time of acceleration $T_{acc}$ that best fits the data. We can see that the effect of the reacceleration is to increase the level of the radio emission, bringing it from a level well below the observed one (about one order of magnitude) up to a level similar to the observed one in relatively short times, $\sim120$ and $\sim450$ Myr for the two values of $\chi$ we have used. Regarding the shape of the spectrum, the model with stronger turbulence (dashed line in Figure \ref{radio_9tt}) produces a spectrum flatter than the one with weaker tubulence (dot-dashed line). The model with stronger turbulence appears to better reproduce the data at high frequencies (especially at 1.4 GHz), whereas the model with lower turbulence is better at low frequencies (30--50 MHz).

\begin{figure}
\centering
\begin{tabular}{c}
\includegraphics[width=\columnwidth]{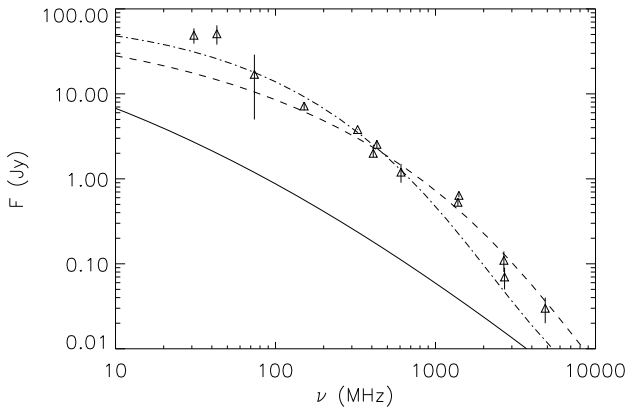}
\end{tabular}
\caption{Spectrum of the radio emission produced for DM models with mass 60 GeV, annihilation final state $b \bar b$, annihilation cross section $\langle \sigma v\rangle=1\times10^{-26}$ cm$^3$ s$^{-1}$, and boosting factor ${\cal B}=30$ for the equilibrium case without reacceleration (solid line), the case with reacceleration with $\chi=1\times10^{-16}$ s$^{-1}$ after $T_{acc}=2.2\times10^8$ yr (dashed line), and the case with reacceleration with $\chi=5\times10^{-17}$ s$^{-1}$ after $T_{acc}=8\times10^8$ yr (dot-dashed line).}
\label{radio_60bb}
\end{figure}

\begin{figure}
\centering
\begin{tabular}{c}
\includegraphics[width=\columnwidth]{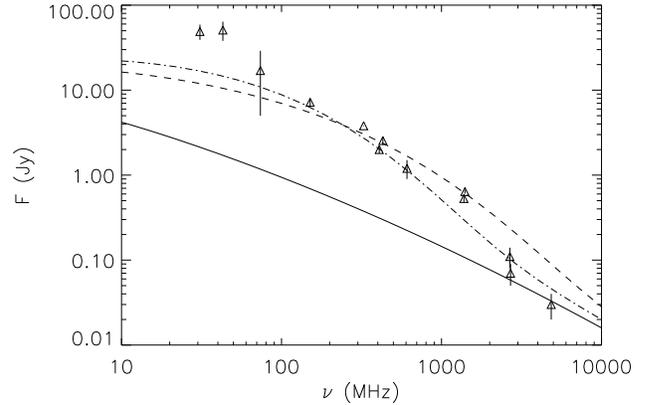}
\end{tabular}
\caption{Spectrum of the radio emission produced for DM models with mass 500 GeV, annihilation final state $W^+W^-$, annihilation cross section $\langle \sigma v\rangle=2\times10^{-25}$ cm$^3$ s$^{-1}$, and boosting factor ${\cal B}=30$ for the equilibrium case without reacceleration (solid line), the case with reacceleration with $\chi=1\times10^{-16}$ s$^{-1}$ after $T_{acc}=2.5\times10^8$ yr (dashed line), and the case with reacceleration with $\chi=5\times10^{-17}$ s$^{-1}$ after $T_{acc}=1\times10^9$ yr (dot-dashed line).}
\label{radio_500ww}
\end{figure}

In Figures \ref{radio_60bb} and \ref{radio_500ww} we show the other two cases of DM particles we have considered. The case with mass 60 GeV is quite similar to the case with mass 9 GeV, with the required total reacceleration time being longer than in the case with 9 GeV. The case with mass 500 GeV requires even longer total acceleration times, and the resulting spectra reproduce less accurately the observed radio halo spectrum.

\begin{figure}
\centering
\begin{tabular}{c}
\includegraphics[width=\columnwidth]{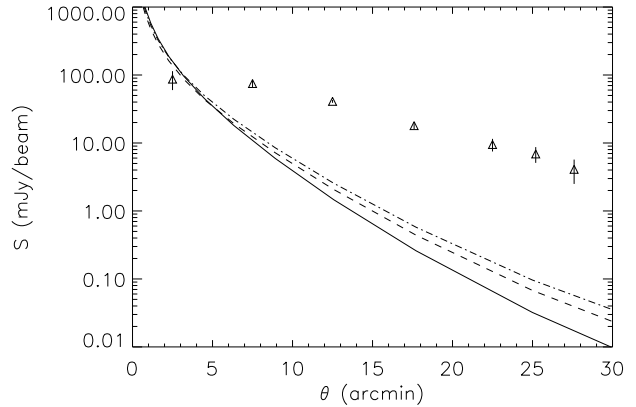}
\end{tabular}
\caption{Azimuthally averaged surface brightness profile at 1.4 GHz for the three cases shown in Figures \ref{radio_9tt}--\ref{radio_500ww} having $\chi=1\times10^{-16}$ s$^{-1}$ for the models with neutralino mass of 9 (solid line), 60 (dashed line), and 500 (dot-dashed line) GeV. Data are from Deiss et al. (1997).}
\label{radio_sb}
\end{figure}

Finally, in Figure \ref{radio_sb} we show the azimuthally averaged surface brightness profile at 1.4 GHz for the three neutralino models we have used in the case with $\chi=1\times10^{-16}$ s$^{-1}$ (the profiles obtained for $\chi=5\times10^{-17}$ s$^{-1}$ result to be very similar to these ones).  It is possible to see that these models do not fit the observed shape of the radio surface brightness; this is a consequence of the quick decrease with radius of the DM distribution (see eq.\ref{dens.dm}), that produces a radio halo profile more peaked than the observed ones, that is typical of DM models (e.g. Colafrancesco et al. 2006).

\section{Discussion}

Among the models we have considered, no one seems to reproduce very well the observed spectrum of the radio halo in the Coma cluster along the whole spectral range. We note, however, that we have considered only a very small subset of all the possible models (in terms of neutralino mass, annihilation final state, intensity and length of the reacceleration). The exploration of a wider set of models with the goal of finding some combination of parameters providing an accurate fitting to the spectrum of the Coma cluster is beyond the goal of this paper, and will be explored in the future, considering also more accurate models for the properties of the turbulent reacceleration, and including also the constraints that can be obtained from the spatial profile of the radio halo surface brigthness. 

It is also important to point out that the spectrum of the Coma radio halo used in this paper has been obtained with a variety of instruments having different sensitivities and systematics, with fluxes that have been obtained integrating in regions with different size depending on the instrument sensitivities, different procedures used to subtract point sources, and different uses of flux calibration scale (see discussion in Brunetti et al. 2013). These effects can produce the scattering present in the data, and make difficult to perform a statistically significant distinction between the fits presented in this paper, at least for the models with mass of 9 and 60 GeV. Therefore we can conclude that the range of parameters explored in this paper allows to match the data, but a rigorous distinction between the models would require a preliminary standardization of the data that is beyond the goals of this paper.

We have explored also the surface brightness profile of the radio emission that is expected, finding that it is more peaked than the observed one. This difference can be explained in different ways: first, as we have mentioned, in order to simplify the calculation we have used a value of the reacceleration parameter $\chi$ constant with the radius, and we have neglected the effect of the spatial diffusion. A reacceleration parameter increasing with the radius can have the effect of broadening the electrons spatial distribution, and consequently the resulting radio surface brightness profile. A broadening of the electrons spatial distribution can be produced also by the spatial diffusion, that anyway should not have a big effect at large distance from the cluster centre for DM halos with mass of the order of a galaxy cluster (Colafrancesco et al. 2006). We note also that in this paper we have considered only the main central DM halo, but it has been found that the observed distribution of DM sub-halos inside the cluster has a spatial shape similar to the one of the radio halo (Brown \& Rudnick 2011), and that considering the sub-halos distribution in DM models produces a radio surface brightness profile more similar to the observed one (Marchegiani \& Colafrancesco 2016). Finally, it is also possible that the contributions of different species of electrons can be important at different distances from the cluster centre, producing in this way a broader radio emission (e.g. Zandanel, Pfrommer \& Prada 2014).

In general, these results show that, other than primary and secondary electrons of hadronic origin, also electrons resulting from DM annihilation can be the seed for the reacceleration due to the turbulences produced following a merging event. According to these results, reacceleration with moderate intensity can provide a radio emission with intensity of the order of the observed one in relatively short times, that are compatible with the times for which turbulent reacceleration is expected to be effective (e.g. Brunetti \& Lazarian 2007), when considering conservative values of the substructures boosting factor, and values of the annihilation cross section in accordance with the upper limits found in other astrophysical experiments. As mentioned, these values have been normalized to the upper limits derived in dwarf galaxies, therefore also the the fluxes that have been calculated are upper limits; if the real values of the annihilation cross section are much lower than the values that have been used, also the corresponding fluxes would be consistently reduced.

As well as models based on reacceleration of electrons produced in hadronic interactions, the models considered in this paper suggest that in clusters that are ``off-state'' (i.e. without turbulent reacceleration effective) a diffuse radio emission about one order of magnitude lower than the observed one should be present because of the electrons continously produced by DM annihilation and in equilibrium with the energy losses. Forthcoming instruments like MeerKAT or SKA should be able to detect this off-state radio emission in relaxed clusters (e.g. Cassano et al. 2012), that in DM models should be directly related to the mass of the cluster.

Where the DM models are different from the hadronic ones is instead in the level of gamma rays they predict: while cosmic ray protons are expected to be reaccelerated by turbulences, this is not expected for DM particles. As a consequence, while 
for models based on reacceleration of electrons of hadronic origin
the level of gamma rays in disturbed galaxy clusters like Coma is expected to be very close to the present upper limits found by Fermi-LAT (e.g. Brunetti et al. 2017), for DM models the level of gamma rays is expected to not be affected by turbulent reacceleration, and to be similar in disturbed and relaxed clusters. In particular, we have calculated that in Coma the gamma ray flux expected for the models we have considered is $F(>0.1 \mbox{ GeV})=1.7\times10^{-11}$, $1.4\times10^{-11}$, and $3.8\times10^{-12}$ cm$^{-2}$ s$^{-1}$ for the 9, 60, and 500 GeV model, respectively, while the upper limit in the same band derived with Fermi-LAT is $4.2\times10^{-9}$ cm$^{-2}$ s$^{-1}$ (Ackermann et al. 2016), i.e. more than two orders of magnitude higher. 
Therefore the detection, or a non-detection, of a gamma ray flux at a level just below the present upper limit values, can point towards a hadronic or a DM origin respectively for the seed electrons that are reaccelerated by turbulences.

Models based on reacceleration of primary electrons (for example accelerated by shocks or injected by AGNs, see e.g. Brunetti et al. 2001) also predict a level of gamma rays much weaker than the hadronic models. A possible way to discriminate between primary or DM-produced electrons in the case of low gamma ray emission might be obtained by looking at the spatial distribution of the radio emission: since the diffusion length of electrons is much smaller compared to the typical size of a radio halo (e.g. Blasi \& Colafrancesco 1999), also after the reacceleration phase the electrons are not expected to move far from the acceleration site. Therefore, for primary electrons we should expect a more irregular spatial distribution of the radio emission, with peaks located close to the acceleration regions (see e.g. Miniati et al. 2001 for the radio emission resulting in the case of a shock front crossing a cluster), whereas for DM-produced electrons we should expect a more regular distribution of the radio emission, with possible sub-peaks located in correspondence of DM sub-peaks (e.g. Marchegiani \& Colafrancesco 2016).

\section{Conclusions}

In this paper we have explored the role of turbulent reacceleration on electrons produced by DM annihilation in the Coma cluster. We have found that, if the annihilation cross section is of the order of the upper limits derived with gamma rays observations of dwarf galaxies, for a moderate level of reacceleration the observed level of the flux of the Coma radio halo can be reached in relatively short times, that are compatible with the expected duration of the reacceleration phase, even using conservative values of the substructures boosting factor. 
Therefore, we conclude that electrons produced by DM annihilation can be possible seeds for the reacceleration activity, as well as primary or secondary electrons of hadronic origin.

For the limited set of reacceleration parameter values and neutralino models used in this paper, 
we have found that the range of parameters that has been adopted can allow to match the data, even if the scattering present in the data does not allow to distinguish between the fits obtained with different models.
Wider ranges of DM models, as well more detailed models for the reacceleration, including also the spatial properties, and possibly a specific standardization of available data, will be considered in future papers.

We also found that the level of diffuse radio emission produced by DM-originated electrons without reacceleration should be about one order of magnitude below the observed flux. Therefore, a diffuse radio emission in other nearby clusters with mass similar to Coma but in a relaxed state should be in principle observable with next-generation of radio instruments like MeerKAT or SKA. 

Finally, a possible way to discriminate between models based on the reacceleration of electrons of DM or hadronic origin is given by the gamma ray emission produced in the cluster, that should be close to the present upper limits in the case of hadronic electrons, and much lower in the case of DM-originated electrons.
The spatial distribution of the radio emission should instead help to distinguish between models based on reacceleration of primary or DM-produced electrons, because in the first case the radio emission should have a spatial distribution more irregular than in the second case.


\section*{Acknowledgments}
This work is based on the research supported by the South African Research Chairs Initiative
of the Department of Science and Technology and National Research Foundation of South
Africa (Grant No 77948). PM acknowledges support from the Department of Science and Technology/National Research Foundation
(DST/NRF) Square Kilometre Array (SKA) post-graduate bursary initiative under the same Grant.
I thank the Reviewer for useful comments and suggestions that helped to improve the quality of the paper.



\bsp

\label{lastpage}

\end{document}